# Predictive autoencoder-transformer model of Cu oxidation state from EELS and XAS spectra

Brian Lee[1], Linna Qiao[2], Samuel Gleason[3,4], Guangwen Zhou[2], Xiaohui Qu[1], Judith Yang,[1,*] Jim Ciston[3,*], Deyu Lu[1,*]

[1]Center for Functional Nanomaterials, Brookhaven National Lab, Upton, NY, United States
[2]Materials Science and Engineering, Binghamton University, Binghamton, NY 13902-6000, USA
[3]National Center for Electron Microscopy Facility, Molecular Foundry, Lawrence Berkeley National Laboratory, Berkeley, CA, USA
[4]Department of Chemistry, University of California, Berkeley, CA, USA

## Abstract

X-ray absorption spectroscopy (XAS) and electron energy-loss spectroscopy (EELS) produce detailed information about oxidation state, bonding, and coordination, making them essential for quantitative studies of redox and structure in functional materials. However, high-throughput quantitative analysis of these spectra, especially for mixed valence materials, remains challenging as diverse experimental conditions introduce noise, misalignment, broadening of the spectral features. We address this challenge by training a machine learning model consisting of an autoencoder to standardize the spectra and a transformer model to predict both Cu oxidation state and Bader charge directly from L-edge spectra. The model is trained on a large dataset of FEFF-simulated spectra and evaluates model performance on both simulated and experimental data. The results of the machine learning model exhibit accurate prediction across the domains of simulated and experimental XAS as well as experimental EELS. These advances enable future quantitative analysis of Cu redox processes under in situ and operando conditions.

# 1. Introduction

The redox state of copper is a critical parameter that governs the performance of materials in fields ranging from electrocatalysis[1, 2], electronics,[3] and batteries[4]. Core-level spectroscopy techniques, such as Electron Energy Loss Spectroscopy (EELS) and X-ray Absorption Spectroscopy (XAS), are uniquely suited for probing this dynamic chemistry at atomic and nanoscale resolutions, respectively. The L-edge spectrum, in particular, provides a sensitive fingerprint of the local 3d electronic structure, directly encoding the Cu oxidation state (e.g., Cu(0), Cu(I), Cu(II)) and its coordination environment.[5, 6] The ability to deploy these techniques in-situ and operando offers a powerful pathway to track these chemical transformations in real-time. The dynamic evolution of chemical systems in the experiment produces vast, complex spectral datasets that, if properly interpreted, can reveal detailed reaction mechanisms. Therefore, the development of robust, quantitative analysis methods to interpret these spectra near real-time is essential for accelerating materials discovery.

Extracting quantitative chemical information from these complex spectra, however, remains a significant challenge.[7, 8] Traditionally, analysis relies on matching spectra to known experimental or simulated standards, or by using linear combination fitting to decompose a mixed-valence spectrum.[9-11] As noted by Gleason et al.[12], this process is highly sensitive to instrumental variations, requiring new, high-quality standards for each experiment and struggles with novel or unexpected chemical environments. To overcome this challenge, machine learning (ML) has emerged as a promising approach to automate and standardize spectral interpretation. Recent works[12-15] have successfully applied ML models to predict chemical properties like Bader charge, coordination number, and oxidation state from large libraries of simulated spectra.

A critical gap, however, remains in transferring these ML models from the clean domain of simulated data to the noisy domain of experimental data. Random forest (RF) models[12, 13] trained on simulated spectra often see their accuracy significantly reduced when applied to experimental data. This simulation-to-experiment gap stems from two primary limitations. First, experimental spectra are subject to instrumental broadening, calibration-dependent energy shifts, and complex noise characteristics not fully captured by standard data augmentation. Second, RF models are architecturally highly sensitive to these variations because they learn to rely on features at absolute energy positions rather than the underlying spectral shape.

In this paper, we endeavor to overcome these limitations by designing a ML model whose architecture is inherently more robust to such experimental variations. We begin with a training dataset, which consists of simulated spectra from Materials Project dataset[16] and additional simulations described in Gleason et al.[12] We augment this simulated data using various data augmentation strategies that include adding Poisson noise, applying random energy shifts, and generating linear mixtures of spectra to mimic mixed-valence states. Second, we develop a ML model with autoencoder-transformer architecture. The autoencoder first learns to denoise the spectra and project them into a low-dimensional latent space. The transformer's self-attention mechanism then learns the contextual features of the spectra, such as peak shapes, intensity ratios, and the energy separation between the $L_3$ and $L_2$ edges. We train this model to output both the oxidation state and the Bader charge, giving a comprehensive view of the electronic structure and the partial charge of the material. The results show that our model performs exceptionally well on multiple domains of simulated as well as on experimental spectra, as measured by the $R^2$ coefficient of determination or mean squared error with the ground truth. This work provides a ML analysis pipeline capable of bridging the simulation-to-experiment gap, laying the foundation for reliable, real-time quantitative analysis of in-situ and operando experiments.

## 2. Methods

**2.1 Machine learning model architecture**

The general schematic for model architecture and training procedure is depicted in Fig. 1. To increase the transferability of ML model to experimental data, we develop a two-step model consisting of an autoencoder (AE) to denoise the Cu L-edge spectra and a transformer (TF) to predict the oxidation state and Bader charge of Cu given its spectra and the latent space from the AE. The AE is a fully convolutional architecture designed for 1D sequential data, consisting of an encoder and a decoder.[17] The encoder down-samples the spectrum to extract key features in a reduced dimensional representation, namely the latent vector. It consists of three sequential convolutional blocks. Each block contains a 1D convolution layer with kernel size of 3, padding of 1, and stride of 2. The convolution layers are followed by a group normalization layer and Gaussian Error Linear Unit (GELU)[18] activation function. The stride of 2 halves the sequence length each while the channel depth is increased as 1→64→128→256. After the final convolutional block, the feature map is flattened and passed through a single fully connected layer to produce final latent vector $z$. We choose the latent vector dimension to be 16 based on hyperparameter testing results of 8, 16, 32 that is not presented in the paper.

The decoder reconstructs the full-length denoised spectrum from the latent vector. The latent vector $z$ is passed through a fully connected layer to restore the flattened map's original dimensions. This tensor is then passed through a series of three up-sampling blocks that consists of an up-sampling layer and a 1D convolution layer (kernel size of 3, padding of 1, stride of 1), a group normalization layer, and a GELU activation function. The decoder channel dimensions change as 256→128→64. The final up-sampling layer maps the 64 channels into a single-channel output spectrum. The AE model yields this output spectrum as well as the latent vector $z$ for TF input.

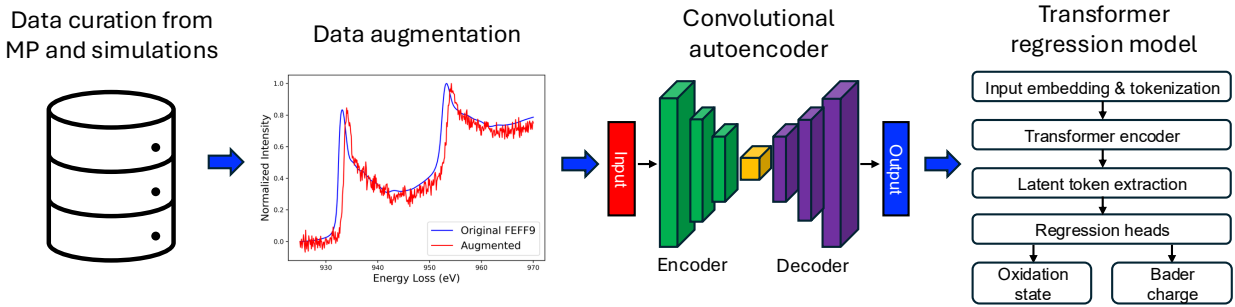

*Figure 1 ML model training procedure and architecture.* *Dataset is obtained from the Materials Project and simulations as described by Gleason et al.[12] This dataset is augmented by adding Poisson noise and misalignment. A 1D convolutional autoencoder is trained to denoise the augmented dataset. A transformer regressor utilizes the original and denoised spectra as well as the latent space from autoencoder to output the oxidation state and the Bader charge.*

A transformer regression model[19, 20] is used to predict the final oxidation state and Bader charge. The inputs for the TF are the original spectral data (input to the AE), denoised output from AE, and the latent vector $z$. The dimensions of the spectra are 451, as described in section 2.3. These original and denoised sequences are tokenized by passing them through a linear layer with an input feature size of 1 and an output feature size of 128 to project them into dimensions of (451, 128). The latent vector is treated as a single token (analogous to [CLS] token[20]) and is passed through a linear layer with input size $z$ and output size 128 to project them into (1,128). These input tokens are concatenated into a single sequence of length 903 (1+451+451). A single, learnable positional embedding with shape (903, 128) is then added to this sequence to provide the model with positional information.

The sequence of 903 tokens is processed by a transformer encoder[19] stack, which allows the model to find complex relationships and contextual features within and between the original and denoised spectra. The encoder consists of 4 layers of transformer encoder modules. Each layer has a model dimension of 128, 4 attention heads, and uses the GELU activation function.

To obtain a final prediction, we use the output of the very first token in the processed sequence (the token corresponding to the AE's latent vector). This 128-dimensional output vector serves as the pooled or aggregate representation of the entire spectrum. This single vector is then fed into two independent regression heads, one for oxidation state and one for Bader charge. Each head consists of a 2-layer linear neural network (128→64→1) with a GELU activation function between them that regresses the final scalar value.

**2.2 Data curation and model training**

The training datasets were generated following the procedure outlined by Gleason et al.[12] The simulated FEFF9[21] spectra of Cu containing materials were curated from the Materials Project[16] with an additional set of 2000 simulations as described in that work. From this dataset, we selected the materials with queryable electron density data from the Materials Project and computed the Bader charge using the Bader analysis code developed by Henkelman et al.[22-24] Additional spectra were generated by linearly mixing individual spectra. All spectra were smoothened through Savitzky-Golay filters[25] and then linearly interpolated onto a uniform grid from 925 eV to 970 eV with 0.1 eV resolution. Finally, the spectra were min-max normalized to be from 0 to 1 in intensity. In Gleason et al., cumulative integrals of the spectra were taken before utilizing the random forest models. While such pre-processing mitigates the effect of misalignment or noise, it also removes many characteristic shapes of the spectra. As we expect our AE-TF model to be more robust against such effects, we retain the original spectral shape without taking the cumulative integrals. To augment the spectra, we added Poisson noise using a random noise level of 0.2. The spectra were also randomly shifted in energy with maximum shift of ±0.5 eV.

## 2.3 Experimental Electron Energy Loss Spectroscopy

Experimental EELS spectra for Cu, $Cu_2O$, and CuO were collected in-lab according to following procedure. For Cu and $Cu_2O$, Single-crystal Cu(100) thin films (~50 nm thick) were deposited onto NaCl(100) substrates by electron-beam evaporation at ~300 °C. After flotation transfer in deionized water, the films were mounted onto Cu mesh TEM grids. All in situ experiments were performed using an FEI Titan 80–300 environmental TEM. EELS data were acquired with an energy dispersion of 0.5 eV/channel, a convergence semi-angle of ~10 mrad, and a collection semi-angle of ~20-25 mrad.

The in situ TEM experiments were carried out in three sequential stages. (1) Native oxide removal: Samples were annealed in flowing $H_2$ (~0.6 Pa) at ~600 °C, during which faceted holes developed within the Cu films. (2) Controlled oxidation: The gas environment was then switched to $O_2$ (~0.3 Pa) at ~300 °C to induce the nucleation and growth of $Cu_2O$ islands. EELS spectra of metallic Cu ($Cu^0$), $Cu_2O$ ($Cu^{1+}$) were obtained during this stage. (3) Reproducibility check: The experiments were repeated on ~50 nm-thick polycrystalline Cu films to ensure the generality of the results.

Reference spectra for CuO were obtained from CuO nanowires synthesized by oxidizing polycrystalline Cu foil (99.99% purity, Sigma–Aldrich) at 450 °C for 2 h under 200 Torr of $O_2$ to grow CuO nanowires[26]. The resulting CuO nanowires were mechanically scraped from the foil, transferred onto TEM grids, and the EELS spectra were collected under the same acquisition conditions as those used for $Cu^0$ and $Cu^{1+}$.

To improve the reliability of oxidation-state classification, all experimental EELS spectra were preprocessed using a standard deconvolution procedure. A power-law background, $I(E) = AE^{-r}$, was first fitted in the pre-edge region and subtracted[27]. Zero-loss peak (ZLP) spectra collected under identical optical conditions were then used as the point-spread function for Fourier-ratio deconvolution, following established EELS deconvolution methodology[28]. The resulting single-scattering distribution was smoothed using a low-pass Fourier filter to suppress high-frequency noise while preserving near-edge fine structure (ELNES), which is essential for distinguishing Cu oxidation states[29]. All spectra were subsequently aligned to the ZLP maximum and normalized to the post-edge intensity prior to further analysis. We find this preprocessing is crucial to model performance for experimental EELS spectra, particularly for data acquired without monochromation.

## 3. Results

### 3.1 Prediction on simulated spectra

The primary objective of this work is to develop a transformer (TF) model capable of accurately predicting the oxidation state and Bader charge of copper from experimental XAS and EELS spectra. As an initial validation, we first evaluated the model's performance on the simulated XAS dataset generated using the FEFF9 code.[30] Figure 2 presents parity plots comparing the model's predictions to the ground-truth values. On the training dataset (Fig. 2a, d), the model shows excellent fidelity, achieving $R^2$ scores of 0.989 for oxidation state and 0.981 for Bader charge. More importantly, the model demonstrates strong generalization to the test dataset (Fig. 2b, e) with similar $R^2$ values as the training datasets.

To further probe the model's capability, we assessed its ability to resolve mixed-valence states, by linearly combining the spectra from metallic Cu, $Cu_2O$, and CuO. These mixed states, which are not explicitly present in the base Materials Project dataset, are highly relevant to chemically heterogeneous environments such as interfacial oxidation. The mixing was performed at a finer resolution than the augmentation used during training, providing a more rigorous test of the model's interpolation capabilities. As shown in Figures 2c and 2f, the TF model accurately predicts the expected values for both oxidation state (MSE = 0.03) and Bader charge (MSE = 0.025), confirming its ability to interpret mixed-valence inputs.

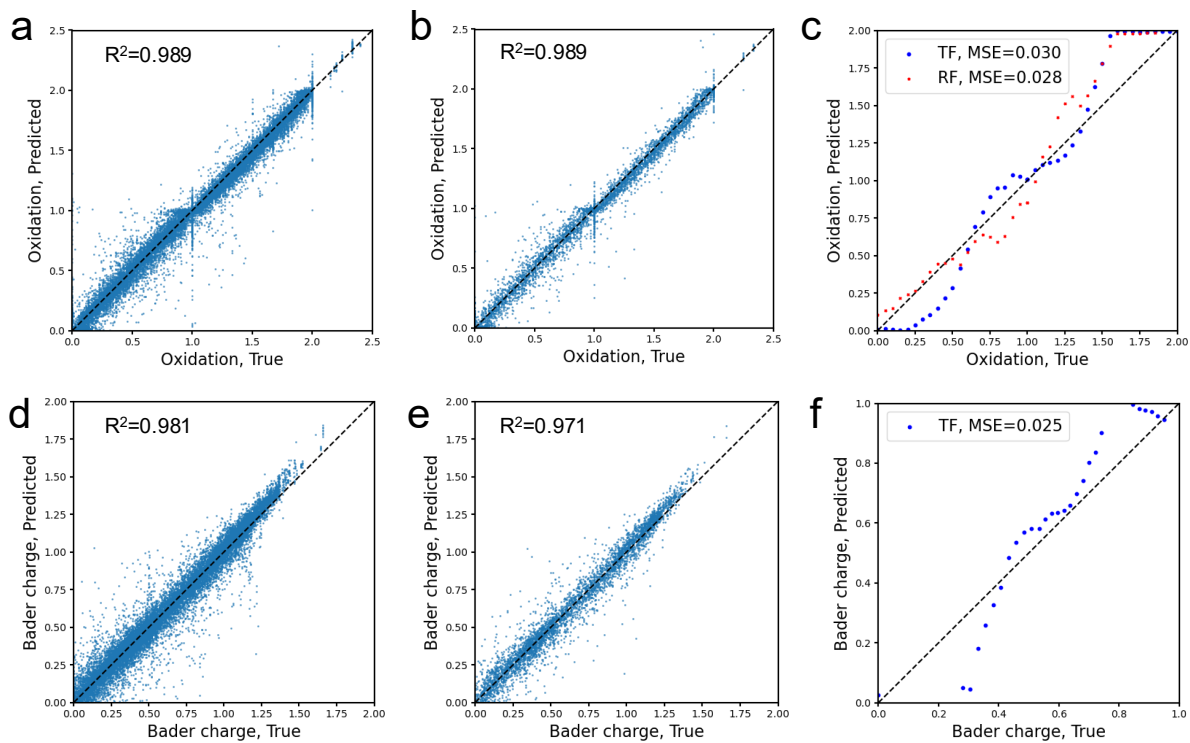

*Figure 2. Model performance on simulated XAS spectra. Parity plots of (a-c) oxidation states and (d-f) Bader charge. (a, d) correspond to the training set and (b, e) test set of all spectra. (c, f) Parity plot of oxidation states obtained from linear mixture of Cu, $Cu_2O$, and CuO spectra.*

To benchmark the performance of the TF model, we also evaluated the random forest (RF) model on the simulated dataset. Results (Fig. 2c and Fig. S1 in the Supplementary Information) show that the RF model is similarly accurate as the TF model for all cases. The strong performance of both ML models on the test datasets demonstrate that the simulated datasets are internally consistent and contain distinct features that the models can utilize to infer the chemical state of Cu.

**3.2 Prediction on experimental XAS and EELS spectra**

We next evaluated the model performances on experimental XAS and EELS spectra were evaluated. All experimental spectra were preprocessed with Savitzky-Golay filter and linear interpolation in the same manner as the simulated spectra that was described in Section 2.2. Fig. 3a shows various XAS spectra of Cu containing materials obtained from literature, with the parity plots shown in Fig. 3b for the TF and RF models on these spectra. The results show that the TF model is highly accurate for these spectra with MSE of 0.001. However, the RF model is less accurate, notably with the Cu (True=0, Prediction=0.327), $Cu_2O$ (True=1, Prediction=0.498), and $CuFeS_2$ (True=1, Prediction=1.591).

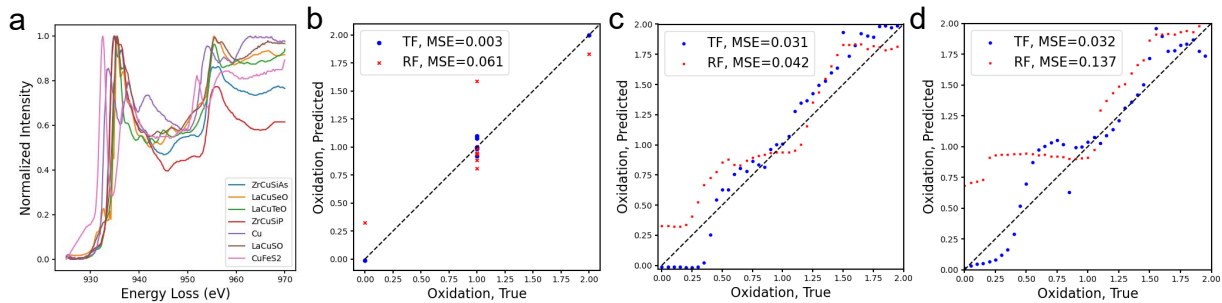

*Figure 3. Model performance on experimental spectra.* (a) XAS spectra of various Cu containing materials from literature[31-34]. (b) TF and RF parity plots of literature spectra. (c,d) Parity plot of linearly mixed Cu, $Cu_2O$, CuO spectra from literature XAS (c) and from in-lab EELS data (d).

We further evaluated the model performance on linear mixtures of Cu, $Cu_2O$, and CuO using literature XAS spectra (Fig. 3c) and in-lab experimental EELS spectra (Fig. 3d). For both modalities, the RF model systematically overestimates the oxidation states of Cu. On the other hand, the TF model did not show the elevated value for metallic Cu and exhibited MSE around 0.03.

Several factors may contribute to this performance gap between the two models. First, Gleason et al.[12] have shown that RF models are much less accurate on the experimental data, especially overestimating the oxidation state of experimental Cu metal, compared to the simulated dataset it was trained on. Second, experimental spectra are subject to various noises, energy shifts, and broadening effects that depend on the instruments and samples. As many of the experimental spectra are not from the same source, the model needs to be robust against such effects. For example, Fig. 3a shows individual plots of some of the experimental spectra. While we utilize Savitzky-Golay filter and linear interpolation to smoothen the spectra before using them as input to the RF or to the autoencoder-TF models, this data pre-processing procedure is not sufficient to remove all noise. While we have trained RF on augmented dataset, the augmentation procedure cannot exhaustively capture the full range of noise characteristics encountered experimentally. The autoencoder-TF pipeline may mitigate such noises from experimental spectra and lead to its outperformance. Overall, the autoencoder-TF model's performance across domains of simulation, experimental XAS, and experimental EELS enables unified analysis pipeline across various modalities.

### 3.3 Model robustness on misaligned data

To mimic common calibration uncertainties in beamline and microscope setups, we applied energy shifts (up to ±2 eV) to experimental XAS spectra from nine compounds depicted in Fig. 3a. This range exceeds the shift for the data augmentation scheme (up to ±0.5 eV) for the simulated spectra used to train the RF and TF. The predictions for the RF and TF models are shown in Fig. 4. In general, the models display somewhat stable predictions for the range of shift (±0.5 eV) observed in the augmented simulated datasets, while showing tendency to significantly deviate from the correct prediction at shift range > 1.5 eV. The RF model displays consistent trend of overestimating the oxidation states for negative shifts and underestimating them for positive shifts. On the other hand, the TF model seems to show regions with spiking errors (see $Cu_2O$ and LaCuSO).

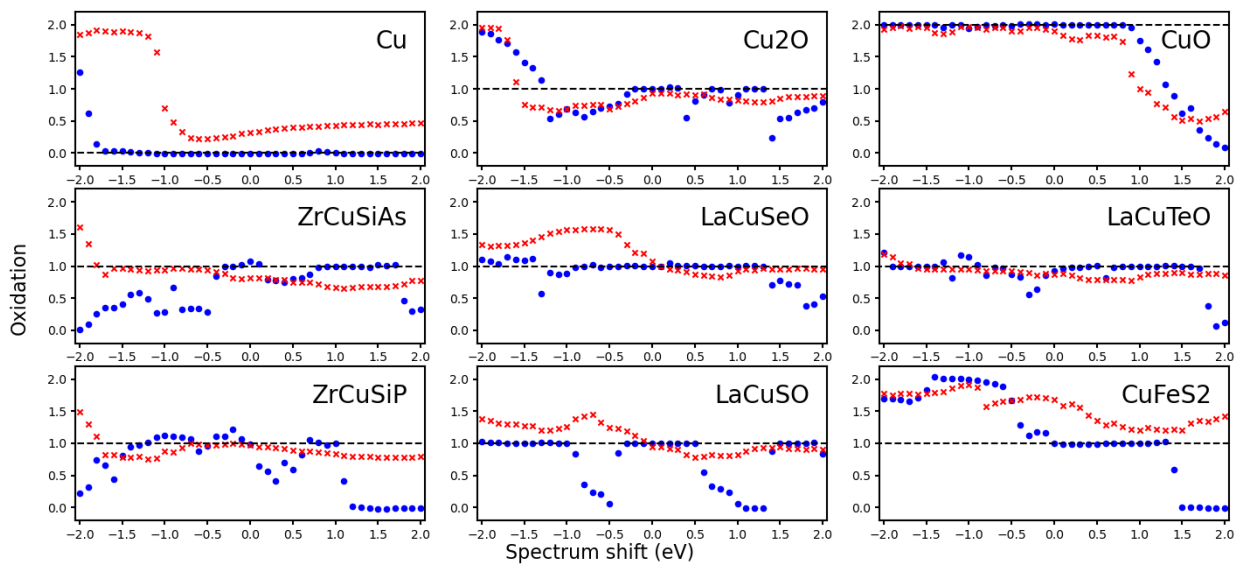

*Figure 4.* Prediction on spectrum shifted dataset. (a) Performance of TF (blue) and RF (red) on nine additional XAS spectra from literature that were shifted $\pm 2eV$. Black lines represent the correct oxidation state.

### 3.4 Experimental mixed valence spectra prediction

We further evaluated the model's capability by testing its prediction on experimental mixed valence spectra from literature,[35] as shown in Fig. 5. The spectra were acquired from a $Cu_2O$ thin film deposited on an n-type silicon substrate that are partially oxidized by post-deposition annealing. The spectra were obtained from depth-dependent regions representing a CuO layer, a mixed $CuO/Cu_2O$ interface, and the underlying bulk $Cu_2O$ phase. While the exact oxidation state of the interfacial region is not known, our model predicts the correct oxidation states of $Cu_2O$ (1.01) and CuO (1.998) spectra, and yields an intermediate oxidation for the mixed interface (1.15). This ability to predict an intermediate value for the interface is encouraging for the model's potential for analyzing Cu redox reactions.

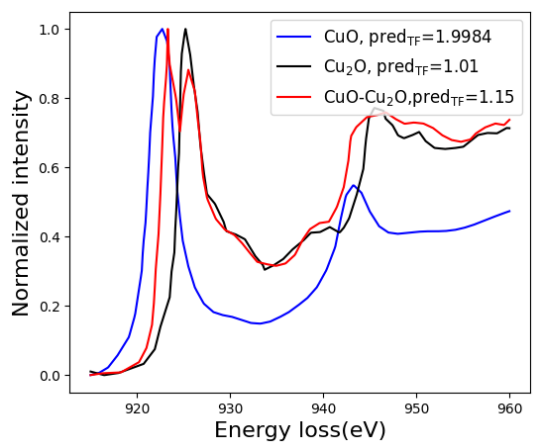

*Figure 5.* Prediction on CuO, $Cu_2O$, and mixed valence spectra extracted from Figure 3 of Lindberg et al[35] using WebPlotDigitizer.[36] As the CuO spectra in the figure is partially obstructed, we use the CuO spectra from the Fig. 3 inset of the paper, which is from Ngantcha et al.[37]

## 4. Conclusions

In this work, we have developed and validated an autoencoder-transformer (AE-TF) model capable of accurately predicting the copper oxidation state and the Bader charge from L-edge XAS and EELS spectra. This ML architecture was designed to be robust against various noises and energy shifts that may be present when the model is applied to experimental data. The results demonstrated that the model outputs accurate oxidation states for both simulated and experimental spectra. Performance against artificially shifted spectra exhibited nearly constant outputs for many cases, indicating that the TF model is not highly dependent on just the absolute energy values but on relative features of the spectra.

This work establishes a framework that can be extended for additional utilities. For example, the transformer model can incorporate conditional inputs to allow its usage for additional elements. In addition to oxidation state or Bader charge, the model can be trained to predict other local chemical descriptors such as the coordination number or local bonding environment. Another aspect that can be improved is the training dataset, such as incorporating both simulated and experimental spectra to produce multi-fidelity model that has shown promising results in other material science applications. Finally, workflow can be developed to apply the model for real-time integration in experiments as feedback to guide automated experiments.

## Acknowledgements

This work was primarily funded by the U.S. Department of Energy program "4D Camera Distillery: From Massive Electron Microscopy Scattering Data to Useful Information with AI/ML." Work at the Molecular Foundry was supported by the Office of Science, Office of Basic Energy Sciences, of the U.S. Department of Energy under Contract No. DE-AC02-05CH11231. This research used Electron Microscopy and Theory and Computation resources of the Center for Functional Nanomaterials (CFN), which is a U.S. Department of Energy Office of Science User Facility, at Brookhaven National Laboratory under Contract No. DE-SC0012704. The work at Binghamton University was supported by the U.S. Department of Energy, Office of Basic Energy Sciences, Division of Materials Sciences and Engineering under Award No. DE-SC0001135.

## Code and data availability.

The code for the transformer model is available on: https://github.com/leebhbrian/ae_transformer_XASEELS.

## Author declarations

The authors have no conflicts to disclose.

# Supplementary Information: Predictive autoencoder-transformer model of Cu oxidation state from EELS and XAS spectra

Brian Lee[1], Linna Qiao[2], Samuel Gleason[3,4], Guangwen Zhou[2], Xiaohui Qu[1], Judith Yang,[1,*] Jim Ciston[3,*], Deyu Lu[1,*]

[1]Center for Functional Nanomaterials, Brookhaven National Lab, Upton, NY, United States
[2]Materials Science and Engineering, Binghamton University, Binghamton, NY 13902-6000, USA
[3]National Center for Electron Microscopy Facility, Molecular Foundry, Lawrence Berkeley National Laboratory, Berkeley, CA, USA
[4]Department of Chemistry, University of California, Berkeley, CA, USA

## Random forest model result on simulation data

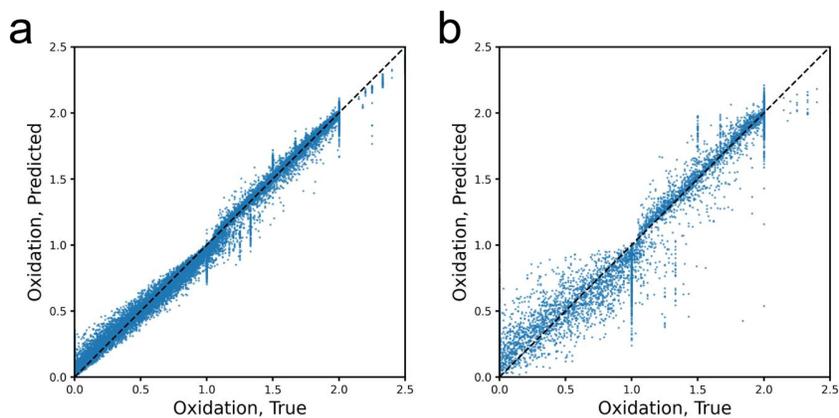

*Figure 1 Random forest model performance on simulated dataset. Parity plots for (a) training and (b) test set of simulated spectra.*